\documentclass{article}

\usepackage{arxiv}
\usepackage[T1]{fontenc}    
\usepackage{url}            
\usepackage{booktabs}       
\usepackage{amsfonts}       
\usepackage{nicefrac}       
\usepackage{microtype}      
\usepackage[noadjust]{cite}

\usepackage{graphicx}
\usepackage{bm}
\usepackage{amsfonts}
\usepackage{color}

\usepackage{amsmath}    
\usepackage{epsfig}
\usepackage{subfigure}  
\usepackage{hyperref}   
\usepackage{bm}
\usepackage{amssymb}

\newcommand{\be}{\begin{equation}}
\newcommand{\ee}{\end{equation}}

\title{TURB-Scalar. A large database of passive scalar fields advected by 2D Navier-Stokes in the turbulent inverse cascade regime}

\author{
  C. Calascibetta \\
  Université Côte d'Azur, Inria, Calisto team \\
  Sophia Antipolis, France\\
  \texttt{chiara.calascibetta@inria.fr} \\
  \And
  L. Biferale \\
  Dept. Physics and INFN\\
  University of Rome Tor Vergata, Italy\\
  \texttt{biferale@roma2.infn.it} \\
   \And
  F. Bonaccorso \\
  Dept. Physics and INFN\\
  University of Rome Tor Vergata, Italy\\
  \texttt{fabio.bonaccorso@roma2.infn.it} \\
     \And
  M. Cencini \\
  Istituto dei Sistemi Complessi, CNR, Rome, Italy\\
  \texttt{massimo.cencini@cnr.it} \\
}

\begin{document}
\maketitle

\begin{abstract}
We introduce TURB-Scalar, an open-access database comprising approximately $400$ uncorrelated snapshots of two-dimensional turbulent velocity and passive scalar fields, obtained from the turbulent inverse cascade regime. These data are generated through Direct Numerical Simulations (DNS) of the advection-diffusion equation for a passive scalar, $\theta$, with resolution $N=4096$. The database serves as a versatile benchmark for the development and testing of both physics-based and data-driven modeling approaches. The scalar field exhibits intermittent statistics with universal anomalous scaling, making TURB-Scalar a valuable resource for studying turbulent transport phenomena. The database is available at \url{http://smart-turb.roma2.infn.it}.
\end{abstract}


\section{Introduction}

Scalar fields transported by turbulent flows arise in a wide range of natural and engineering contexts, including atmospheric dynamics~\cite{pasquill1983atmospheric} and combustion processes~\cite{williams2018combustion}. A particularly rich area of study involves passive scalar fields, which, despite not influencing the flow dynamics, display complex statistical behavior such as intermittency and universal anomalous scaling exponents, largely independent of the injection mechanism~\cite{shraiman2000scalar,falkovich2001particles}.

In this short note, we introduce TURB-Scalar, an open-access database of passive scalar fields advected by two-dimensional turbulent flows. The evolution of the passive scalar field $\theta$ is governed by the advection-diffusion equation:
\begin{equation}
\partial_t \theta + \mathbf{v} \cdot \nabla \theta = \kappa \Delta \theta + f_\theta,
\label{eq:theta}
\end{equation}
where $\kappa$ is the scalar diffusivity, and $f_\theta$ is a large-scale Gaussian forcing~\cite{celani2000universality}. The advecting velocity field $\mathbf{v}$ is incompressible and obeys the 2-dimensional Navier-Stokes equations:
\begin{equation}
\partial_t \mathbf{v} + (\mathbf{v} \cdot \nabla) \mathbf{v} = -\nabla p + \nu \Delta \mathbf{v} - \beta \mathbf{v} + \mathbf{f},
\label{eq:NS}
\end{equation}
where $p$ is the pressure, $\nu$ the kinematic viscosity, $\beta$ a large-scale friction coefficient that prevents the accumulation of energy due to the inverse cascade, and $\mathbf{f}$ is a small-scale Gaussian forcing.

The TURB-Scalar current release includes approximately 400 uncorrelated snapshots of the scalar field $\theta$, the vorticity, the velocity magnitude, and both velocity components $v_x$ and $v_y$. To ensure statistical decorrelation, snapshots are stored at intervals of roughly half of the large-scale eddy turnover times, as discussed in detail in the following sections.

Our aim is to provide a high-resolution, physics-based dataset that serves as a testing ground for a broad range of research communities and methodologies, from data-driven approaches to theoretical and model-based investigations. A first application of this dataset has been presented in~\cite{calascibetta2025hidden}.

The structure of this paper is as follows: in Sec.~2, we provide a brief description of the Direct Numerical Simulations (DNS) used to generate the TURB-Scalar database; in Sec.~3, we discuss the selection criteria for the 400 fields and explain how to access the complete dataset.

\section{Numerical Simulations}

The TURB-Scalar velocity field is evolved via fully resolved DNS of the incompressible 2D Navier-Stokes equations in vorticity form:
\begin{equation}
\partial_t \omega + J(\omega, \psi) = -\nu (-\Delta)^p \omega - \beta \omega + f_\omega,
\label{eq:vorticity}
\end{equation}
where $\omega = \nabla \times \mathbf{v}$ is the vorticity, and $\psi$ is the stream function such that $\mathbf{v} = \nabla^\perp \psi = (\partial_y \psi, -\partial_x \psi)$. The term $J(\omega, \psi)$ denotes the Jacobian determinant, capturing the nonlinear advection. Energy is extracted from the system at large scales via a linear friction term $-\beta \omega$, which prevents accumulation of energy due to the inverse cascade. The characteristic friction scale is given by $L_\beta \sim \epsilon^{1/2} \beta^{-3/2}$, where $\epsilon$ is the energy flux towards the large scales.
The small-scale Gaussian forcing $f_\omega$ has correlation:
$$
\langle f_\omega(\mathbf{x}, t) f_\omega(\mathbf{0}, t') \rangle = \delta(t - t') F_\omega(|\mathbf{x}|/\ell_{f_\omega}),
$$
where $F_\omega(r) = A_\omega \ell_{f_\omega}^2 \exp(-r^2/2)$ ensures compact support around the forcing scale $\ell_{f_\omega}$. As a result, the inertial range dynamics that supports the inverse energy cascade spans the scales $\ell_{f_\omega} \ll \ell \ll L_\beta$. To dissipate enstrophy at small scales, we employ a hyperviscous term of order $p = 8$, consistent with established practices~\cite{boffetta2000inverse}.

The passive scalar field $\theta$ is evolved according to a modified advection-diffusion equation:
\begin{equation}
\partial_t \theta + \mathbf{v} \cdot \nabla \theta = -\kappa (-\Delta)^2 \theta + f_\theta,
\label{eq:thetaDNS}
\end{equation}
where $\kappa$ is a bi-Laplacian diffusivity. The scalar is forced at large scales by a Gaussian forcing $f_\theta$, with correlation:
$$
\langle f_\theta(\mathbf{x}, t) f_\theta(\mathbf{0}, t') \rangle = \delta(t - t') F_\theta(|\mathbf{x}|/L_{f_\theta}),
$$
where $F_\theta$ shares the same functional form as $F_\omega$. The scalar forcing scale $L_{f_\theta}$ is chosen to be slightly smaller than $L_\beta$, in order to avoid overlap with the velocity field's large-scale dissipation range. Moreover, the scalar dissipation scale is set to be larger than the velocity forcing scale, ensuring that scalar dynamics unfold entirely within the inertial range of the turbulent flow. 
The simulations are performed using a standard $2/3$-dealiased pseudospectral method in a doubly periodic domain of size $4096 \times 4096$, parallelized on GPUs using openACC. Time integration is carried out using a second-order Adams-Bashforth scheme. 

\begin{table}[h!]
\centering
\begin{tabular}{cccccc}
\toprule
$N$  & $L$  & $dt$  &  $\nu$  & $\beta$ & $\kappa$  \\
4096 & $2\pi$ & $1.5\times 10^{-5}$ & $1.22\times 10^{-49}$  & $0.03$ & $2.5 \times 10^{-11}$ \\
\midrule
$A_\omega$ & $A_\theta$ & $\ell_{f_\omega}$ & $L_{f_\theta}$ & $\tau_0$ & $\Delta t_{\rm s}$ \\
$0.3125\times 10^{-6}$ & $0.1$ & $\simeq 0.0075$ & $\simeq 1.26$ & $\simeq 14.5$ & $\simeq 0.2\,\tau_0$ \\
\bottomrule \\
\end{tabular}
\caption{Parameters of the DNS: $N$ resolution in each dimension; $L$ physical dimension of the 2-periodic box; $dt$ time step in the DNS integration; $\nu$ kinematic viscosity; $\beta$ large scale friction parameter; $\kappa$ scalar hyper-diffusivity; $A_\omega$ forcing amplitude of the vorticity field; $A_\theta$ forcing amplitude of the scalar field; $\ell_{f_\omega}$ scale of the peak of the vorticity forcing; $L_{f_\theta}$ scale of the peak of scalar forcing; $\tau_0$ large scale eddy turnover time; $\Delta t_{\rm s}$ time interval between two saved configurations in the TURB-Scalar database.}
\label{tab:parameters}
\end{table}
The complete set of parameters used in the simulations are reported in Table~\ref{tab:parameters}.

\section{DataBase Description}
TURB-Scalar database is extracted during the simulation described in the previous section. The total energy evolution is shown in Fig. 1(a). Along this trajectory we have dumped roughly 400 fields equispaced in time. Snapshots are chosen with a temporal separation large enough to decrease correlations in time between two successive data-points. At each time point we dumped in binary format the following fields:
\begin{itemize}
    \item passive scalar
    \item velocity module
    \item $x$ component of the velocity
    \item $y$ component of the velocity
    \item vorticity.
\end{itemize} 
\begin{figure}
    \centering
    \includegraphics[width=1.0\textwidth]{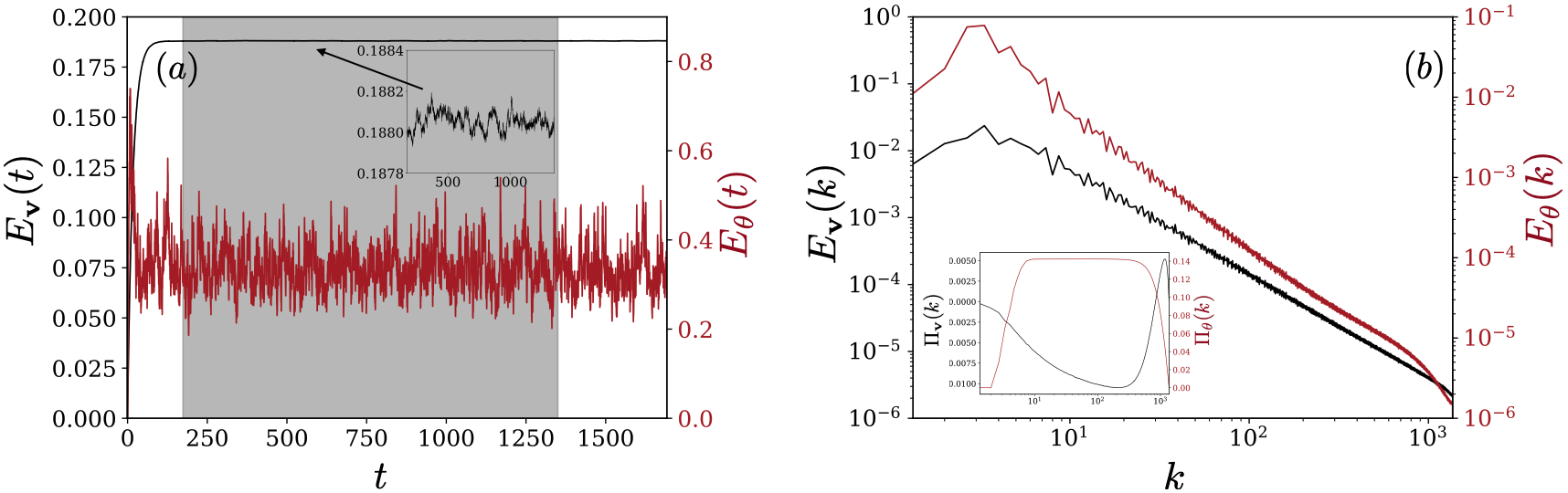}
    \caption{(Panel a) Energy evolution for the turbulent flow (black) and for the passive scalar field (red) generated during the simulation performed to generate the database of $400$ different configurations. The fields are extracted from time $t=174$ up to  $t=1350$ every $\Delta t_{\rm s} \simeq 0.2\, \tau_0$ (gray area). In the inset we show the energy evolution for the velocity $\bold v$ during this time period. (Panel b) Log-log plot of the averaged energy spectrum for the turbulent flow (black) and for the passive scalar field (red). Inset: energy flux $\Pi(k)$.}
    \label{fig:simulation}
\end{figure}
The energy spectra is shown in Fig. 1(b), while in Fig.2 we show the probability density function (PDF) of normalized velocity and scalar increments (panels $a$ and $b$ respectively). The velocity exhibits almost gaussian statistics, while the scalar shows intermittent tails of the PDFs.  
\begin{figure}
    \centering
    \includegraphics[width=1.0\textwidth]{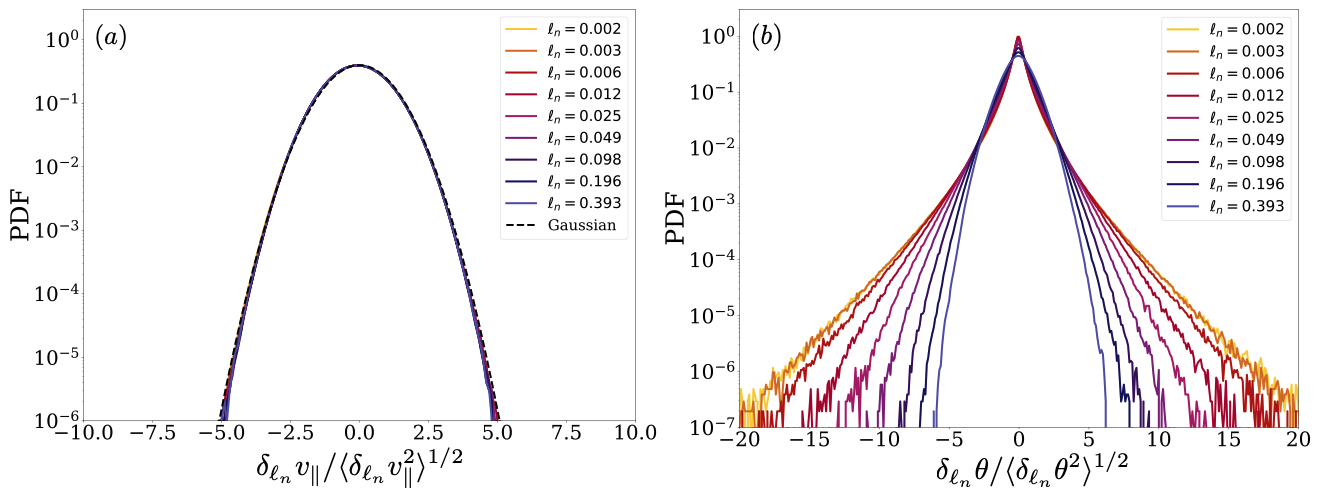}
    \caption{(Panel a) Probability density function (PDF) of normalized longitudinal velocity increments, $\delta_{\ell_n} v_\parallel / \langle \delta_{\ell_n}v^2_\parallel  \rangle^{1/2} $ for different scales $\ell_n = 2\pi/4096 \,2^{n}$. The dashed line represents a gaussian distribution. (Panel b) PDF of normalized scalar increments, $\delta_{\ell_n} \theta_\parallel / \langle \delta_{\ell_n}\theta^2_\parallel  \rangle^{1/2} $. }
    \label{fig:increments}
\end{figure}
The database TURB-Scalar is available for download using the SMART-Turb portal \url{http://smart-turb.roma2.infn.it}. The portal is based on the concept of "Dataset" to aggregate resources related to turbulent simulations, (see other datasets released in the same portal, \cite{biferale2020turb-rot,biferale2023turb-lagr,biferale2024turb-hel,capocci2025turb-mhd,biferale2023turb-smoke}). Details on how to access the data with a few examples can be found on the portal. \\
\noindent Finally, In Fig.~\ref{fig:database1} we show a few examples of the images contained in TURB-Scalar made out of the scalar field. In Figs~\ref{fig:database2}-\ref{fig:database4} we show  for the same snapshots the images generated looking at the velocity magnitude and the two different velocity components.

\begin{figure}
    \centering
    \includegraphics[width=0.8\textwidth]{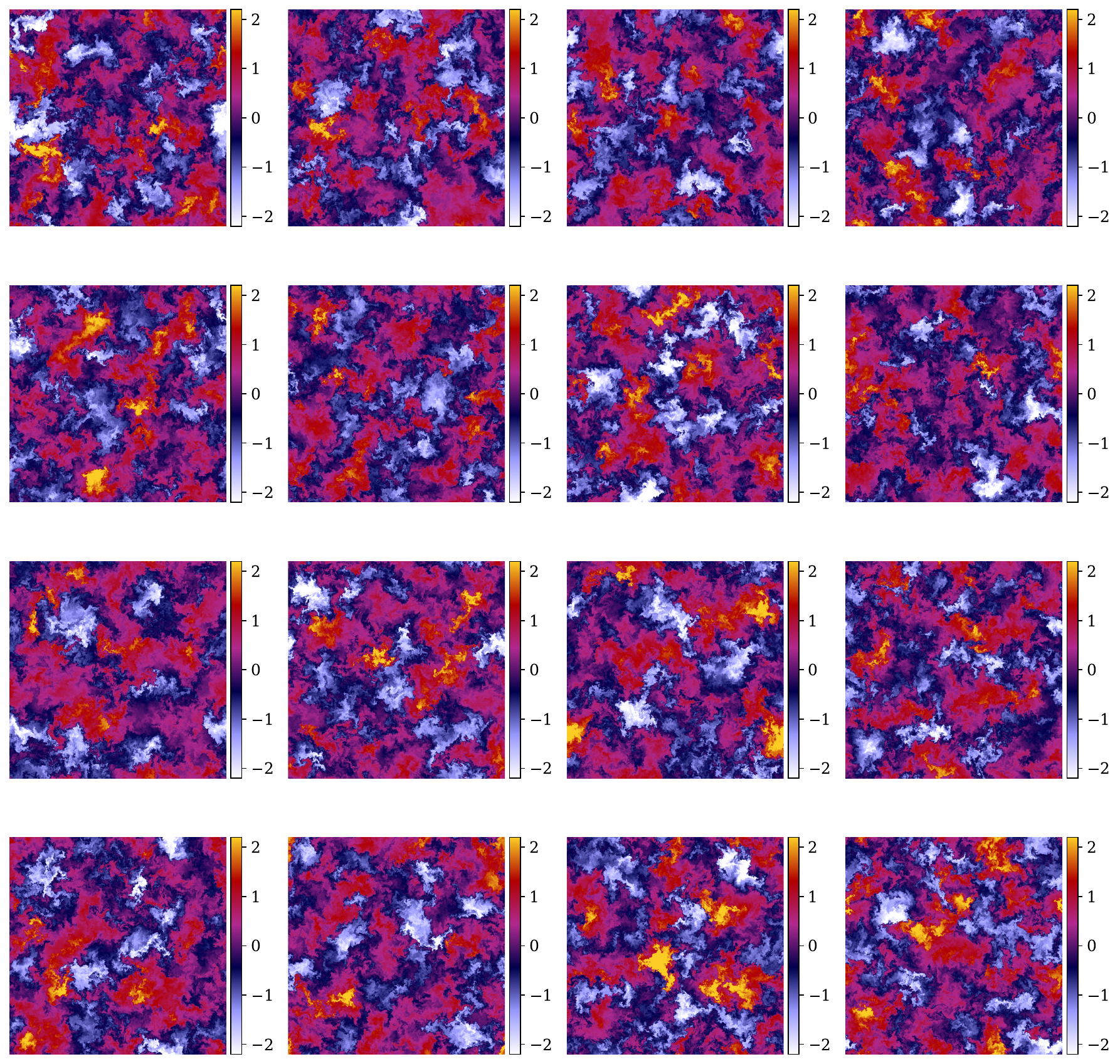}
    \caption{Example of $16$ different $4096\times4096$ passive scalar fields composing the TURB-Scalar database.}
    \label{fig:database1}
\end{figure}
\begin{figure}
    \centering
    \includegraphics[width=0.8\textwidth]{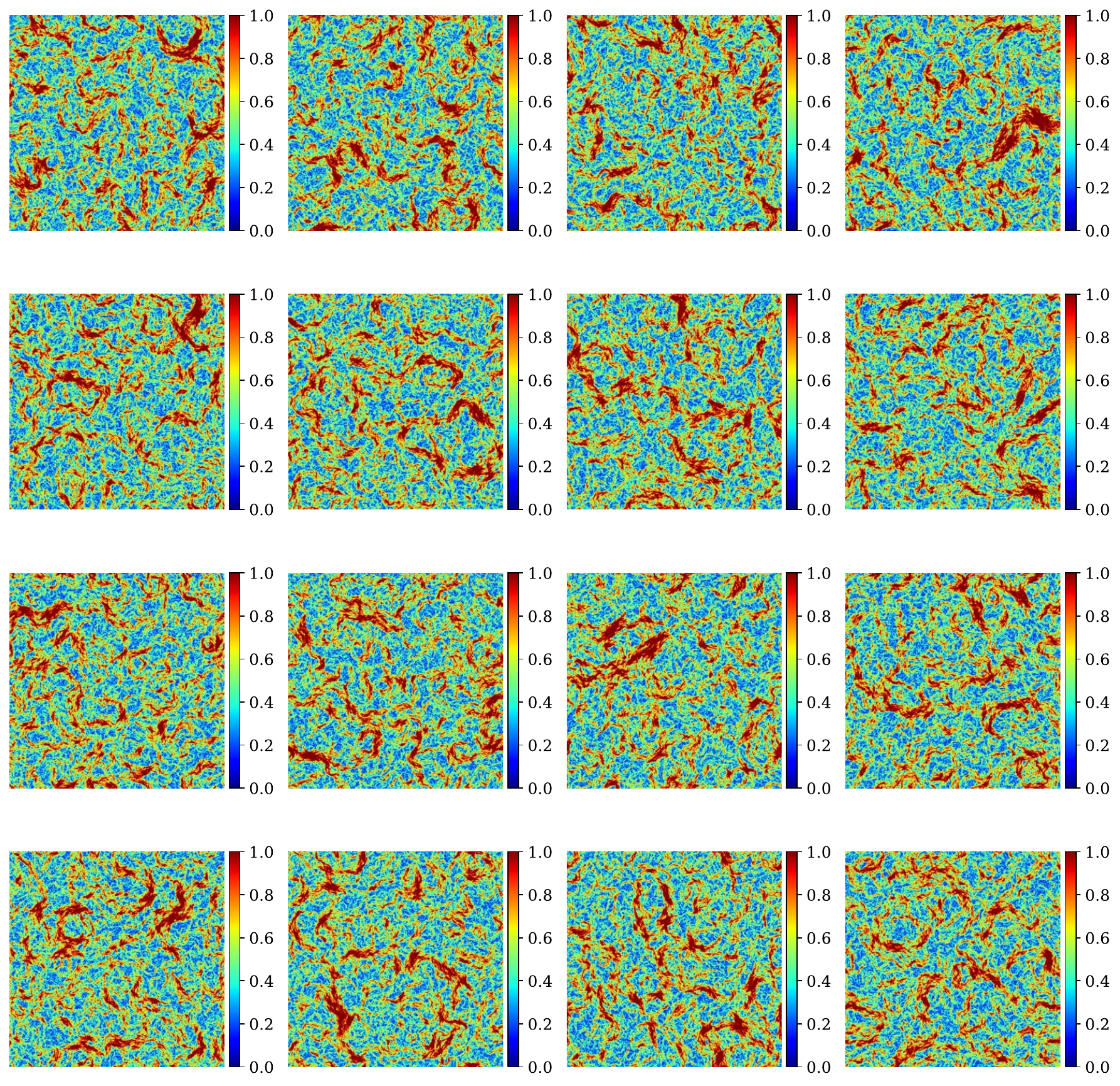}
    \caption{Snapshots at the same times as in Fig.~\ref{fig:database1}, but showing the amplitude of the velocity fields $|\bold v|$.}
    \label{fig:database2}
\end{figure}
\begin{figure}
    \centering
    \includegraphics[width=0.8\textwidth]{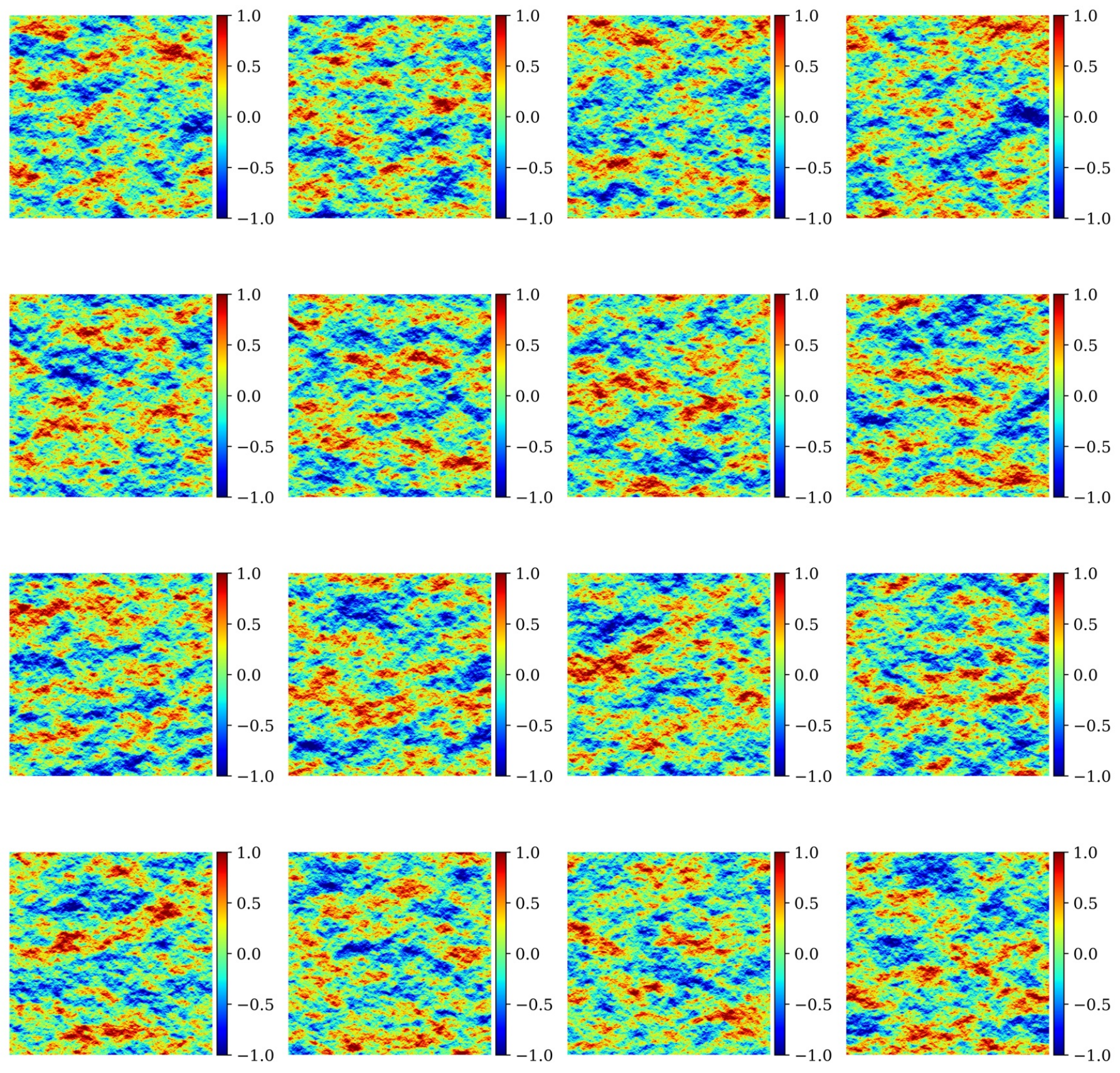}
    \caption{Snapshots at the same times as in Fig.~\ref{fig:database1}, but for the ${v_x}$ component of the velocity field.}
    \label{fig:database3}
\end{figure}
\begin{figure}
    \centering
    \includegraphics[width=0.8\textwidth]{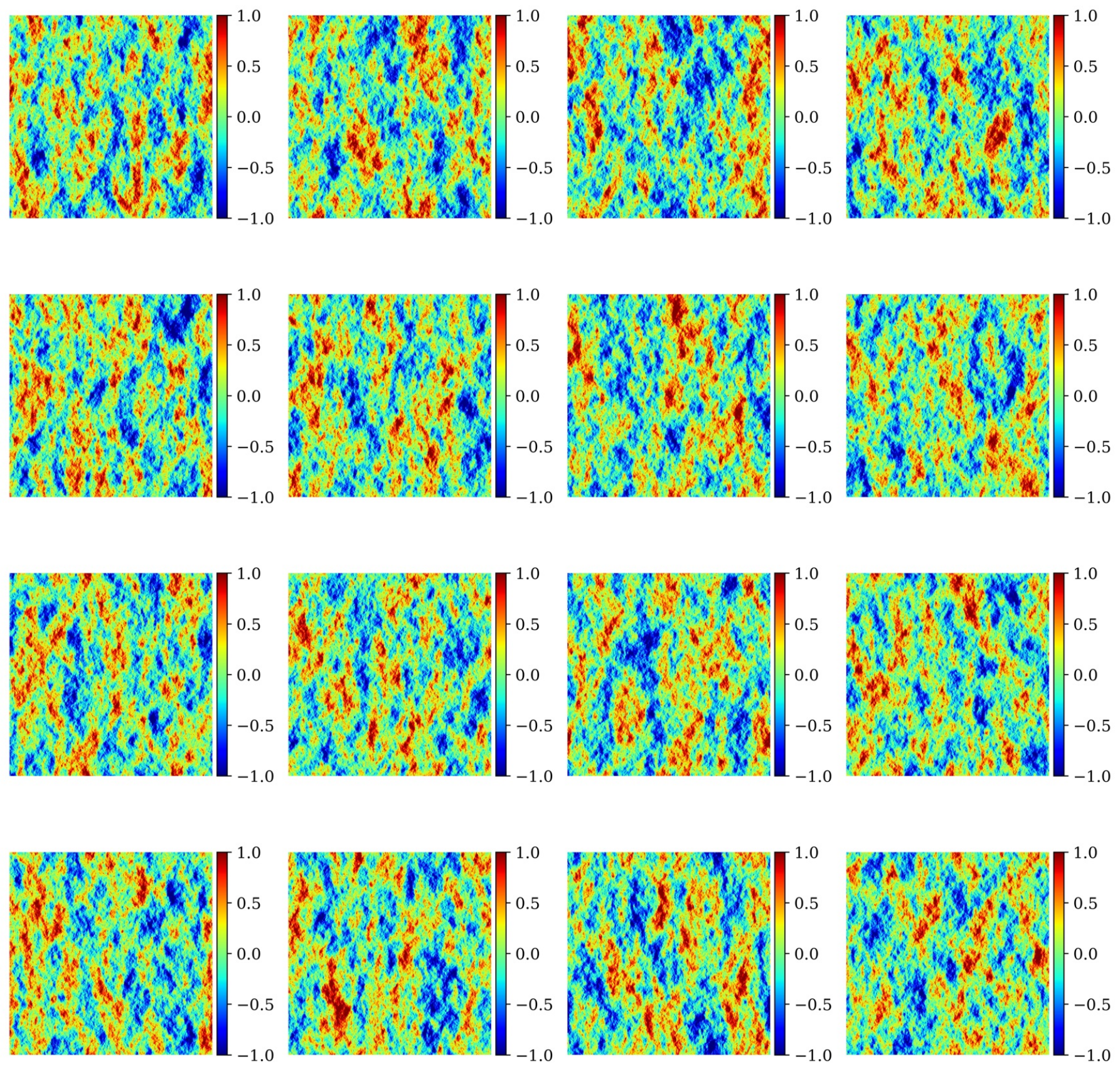}
    \caption{Snapshots at the same times as in Fig.~\ref{fig:database1}, but for the ${v_y}$ component of the velocity field.}
    \label{fig:database4}
\end{figure}

\paragraph*{Acknowledgments}
We acknowledge useful discussions with Guido Boffetta. We also acknowledge
financial support under the National Recovery and Resilience Plan (NRRP), Mission 4, Component 2,
Investment 1.1, Call for tender No. 104 published on 2.2.2022 by the Italian Ministry of University
and Research (MUR), funded by the European Union – NextGenerationEU– Project Title “Equations
informed and data-driven approaches for collective optimal search in complex flows (CO-SEARCH)”,
Contract 202249Z89M. – CUP B53D23003920006 and E53D23001610006. This work was supported
by the European Research Council (ERC) under the European Union’s Horizon 2020 research and
innovation program Smart-TURB (Grant Agreement No. 882340). 

\bibliographystyle{unsrt}
\bibliography{references} 

\end{document}